\def\di{\mbox{d}}
\def\Tr{\mathop{\mbox{Tr}}\,}
\newcommand{\bea}{\begin{eqnarray}}
\newcommand{\be}{\begin{equation}}
\newcommand{\eea}{\end{eqnarray}}
\newcommand{\ee}{\end{equation}}
\newcommand{\nn}{\nonumber}
\documentstyle[bibnorm]{lamuphys}
\begin{document}
\title{Graviton scattering in matrix theory and supergravity}
\author{Marco\,Fabbrichesi}
\institute{INFN, Sezione di Trieste and\\
Scuola Internazionale di Studi Superiori Avanzati\\
via Beirut 4, I-34014, Italy.}
\maketitle
\begin{abstract}
I briefly review recent work on the comparison between
two and three graviton scattering in 
supergravity and matrix theory
\end{abstract}
\section{Motivations}

In the low-energy regime, M-theory is 
$D=11$, $N=1$ supergravity. In the matrix model  
the fundamental degrees of freedom of 
M-theory are 0-branes (that is, Derichlet particles). 
For this model to be a correct
description of M-theory, it must then reproduce supergravity in
the long-distance regime. In particular,
0-brane scattering amplitudes in $D=10$ must reproduce those of
compactified (from $D=11$ down to 10) supergravity,
for which the gravitons carry momentum in the compactified direction.

Such a correspondence between amplitudes in these two 
different-looking theories plays an important role because
it can be computed explicitely.
It has now been succesfully checked in the two- and three-graviton 
scattering amplitudes.

\section{Two-graviton scattering}

The scattering of two graviton carrying momentum in a
compactified direction has been studied several times in the 
literature~\cite{2grav}.
The simplest way to compute it is by means of the effective 
lagrangian~\cite{BBPT}
\be
 L = - p_- \dot x^- = - p_- \frac{\sqrt{1 - h_{--} v^2} -1}{h_{--}}\, ,
\ee
where $h_{--} = f(r)/2 \pi R_{11}$ and 
$f(r) = 2 \kappa^2 M/7 \Omega\, r^7$ for the space-time of the shock
wave generated by the graviton moving with momentum 
$p_- = N_2/R_{11}$. Actually, this is a special case of shock wave
in which the  11-th dimension has been smeared. By expanding in the relative
velocity $v$, we find
\be
L = - p_- \left\{ \frac{v^2}{2} + a_1 \: \frac{v^4}{r^7} + 
a_2 \: 
 \frac{v^6}{r^{14}}  \cdots \right\}\, ,
\ee
where the exact values of the coefficients $a_1$ and $a_2$ are known.

The corresponding amplitude in matrix theory can derived from the
 gauge fixed action, the bosonic part of which reads 
\bea
S &=& \int \di t \: \:\Tr\bigg(\dot a_0^2 + \dot x_i^2 + 
4\,i\,\dot R_k\,[a_0, x_k] 
-[R_k, a_0]^2 - [R_k, x_j]^2\nn\\
&&+2\,i\,\dot x_k\,[a_0, x_k] + 2\,[R_k, a_0][a_0, x_k] 
-2\,[R_k, x_j][x_k, x_j] \nn\\
&&-[a_0,x_k]^2 - \frac{1}{2}[x_k, x_j]^2 \bigg), \label{action}
\eea
where $a_0$ and $x_k$ are hermitian matrices representing the fluctuations 
and $R_k$ is the background. The fermionic and ghost terms must also be 
included in addition to (\ref{action}) but are here omitted for semplicity.

The  units are such that
\be
g_{\mbox{\rm \scriptsize YM}}=\left( R_{11}/
\lambda_{\mbox{\rm \scriptsize P}}^2 \right) ^{3/2}=1 \, ,
\ee
the quantities $R_{11}$, $\lambda_{\mbox{\rm \scriptsize P}}$ 
and $g_{\mbox{\rm \scriptsize YM}}$ 
being the 
compactification radius, the Planck length and the Yang-Mills
coupling, respectively.

The relevant gauge group depends on the process
under stady. It is the rank one (only one independent velocity)
group $SU(2)$ in the two-body scattering.

The corresponding computations at one- and two-loop level
 in matrix theory yield
\be
a_1 = \frac{15}{16} \: \frac{N_1 N_2}{R^3 M^9}
\quad \mbox{(one loop)~\cite{BBPT}}
\ee
and
\be
a_2 = \frac{225}{64} \: \frac{N_1^2 N_2}{R^5 M^{18}}
\quad \mbox{(two loops)~\cite{BC}} \, ,
\ee
in agreement with what found in supergravity.

\section{Three-graviton scattering}

The simplest way to obtain supergravity amplitudes is by means of string
theory. Since it is a tree-level amplitude, it is consistent with
conformal invariance in any
dimensionality, in particular in $D=11$. We consider the {\it bona fide} 
superstring theory (where there is no tachyon) and the scattering amplitude
of three ($11$-dimensional) gravitons, and look at suitable {\it pinching}
limits,
where only intermediate massless states are coupled to the external
gravitons. Those states are themselves $11$-dimensional gravitons.
We then compactify the $10^{\rm th}$ space dimension giving mass
to the external gravitons, which will thus correspond to
$10$-dimensional $D0$-branes. Keeping zero momentum transfer in
the  $10^{\rm th}$ dimension, the intermediate states remain massless
and correspond to the various massless fields of $10$-dimensional
supergravity.

By  considering only the part of the complete amplitude that is
proportional to 
\be
\varepsilon_1 \cdot \varepsilon_1' \: 
\varepsilon_2 \cdot \varepsilon_2' \: \varepsilon_3 \cdot \varepsilon_3' \, ,
\ee
$\varepsilon$ being the external graviton polarization tensor,
we obtain the amplitude $A_6$ for six graviton vertices~\cite{FIR}:
\bea
A_6 & = & \varepsilon_1 \cdot \varepsilon_1' \: 
\varepsilon_2 \cdot \varepsilon_2' \: \varepsilon_3 \cdot \varepsilon_3' \:
\frac{\kappa^4 (\alpha')^3}{4 \pi^3} \int \di ^2 x\: \di ^2 y\: \di z^2 
 |1-y|^{-2 + \alpha' p_2'\cdot p_2} \nn \\
&&\times \: |y|^{\alpha' p_3\cdot p_2'} 
|1-x|^{\alpha' p_2\cdot p_1'} |x|^{\alpha' p_3\cdot p_1'}
 |1-z|^{\alpha' p_3'\cdot p_2} \nn \\
&&\times \:  |z|^{-2 + \alpha' p_3\cdot p_3'}
|z-x|^{\alpha' p_3'\cdot p_1'} |z-y|^{\alpha' p_3'\cdot p_2'}
 |x-y|^{\alpha' p_2'\cdot p_1'} \nn \\
&&\times \: \left\{ \frac{p_3' \cdot p_1' \: p_2' \cdot p_1'}{(y-x)(z-x)} +
\frac{p_3 \cdot p_2' \: p_3' \cdot p_1'}{y(z-x)} -
\frac{p_3' \cdot p_2' \: p_3 \cdot p_1'}{x(z-y)} \right. \nn \\
&& \left. + \frac{p_2' \cdot p_3' \: p_2' \cdot p_1'}{(y-x)(z-y)} +
\frac{p_3' \cdot p_2 \: p_2' \cdot p_1'}{(z-1)(y-z)} \right\} 
\wedge \Biggl\{ c.c. 
\Biggr\} 
\eea 
where $p_i = (E_i, {\bf p}_i-{\bf q}_i /2, M_i)$, 
$p_i' = (-E_i', - {\bf p}_i-{\bf q}_i /2, -M_i)$,
$ p_i^2=0$, $E_i \simeq M_i + ({\bf p}_i-{\bf q}_i /2)^2/2M_i$
and $M_i=N_i/R_{11}$. Moreover, we have that
$\sum_i {\bf q}_i = 0$ and $\sum_i {\bf p}_i \cdot {\bf q}_i = 0$.

In the long-distance regime we are interested in we find that
$A_6 = A_\vee + A_Y$ where
\bea
A_\vee & = & 2 \:
\kappa^4 \: \varepsilon_1 \cdot \varepsilon_1' \: 
\varepsilon_2 \cdot \varepsilon_2' \: \varepsilon_3 \cdot \varepsilon_3' 
\; \frac{1}{{\bf q}_1^2\: {\bf q}_2^2} \nn \\
&& \times \left\{
({\bf p}_3 - {\bf p}_2)^2 \: ({\bf p}_3 - {\bf p}_1)^2  \left[
({\bf p}_2 - {\bf p}_1)^2 -  ({\bf p}_3 - {\bf p}_1)^2 - 
({\bf p}_3 - {\bf p}_2)^2 \right]
  \right. \nn \\
&& -\: ({\bf p}_3 - {\bf p}_2)^2 \: ({\bf p}_3 - {\bf p}_1)^2
\left[ 
({\bf p}_3 - {\bf p}_2)^2 \: \frac{ {\bf q}_2 \cdot  ({\bf p}_3 - {\bf p}_1) }
{ {\bf q}_1 \cdot  ({\bf p}_3 - {\bf p}_1)} \right. \nn \\
&& \left. \left .  + \:
({\bf p}_3 - {\bf p}_1)^2 \: \frac{ {\bf q}_1 \cdot  ({\bf p}_3 - {\bf p}_2) }
{ {\bf q}_2 \cdot  ({\bf p}_3 - {\bf p}_2)} \right] 
 \right\} \: + \:  \mbox{symmetric} 
\eea
and
\bea
A_Y & = & - 2 \: 
\kappa^4 \: \varepsilon_1 \cdot \varepsilon_1' \: 
\varepsilon_2 \cdot \varepsilon_2' \: \varepsilon_3 \cdot \varepsilon_3' 
\; \frac{1}{{\bf q}_1^2\: {\bf q}_2^2\: {\bf q}_3^2} \nn \\
& &  \times \: \Biggl\{  
({\bf p}_2 - {\bf p}_3)^2 
\Bigl[
{\bf q}_3 \cdot ({\bf p}_3 -{\bf p}_1) + 
{\bf q}_2 \cdot ({\bf p}_1 -{\bf p}_2) 
\Bigr] \Biggr. \nn \\ 
&& \quad  + \: ({\bf p}_3 - {\bf p}_1)^2
\Bigl[
{\bf q}_3 \cdot ({\bf p}_2 - {\bf p}_3) +
{\bf q}_1 \cdot ({\bf p}_1 - {\bf p}_2) 
\Bigr]  \nn \\ 
&& \Biggl. \quad  +\:  ({\bf p}_1 - {\bf p}_2)^2
\Bigl[
{\bf q}_2 \cdot ({\bf p}_2 -{\bf p}_3) +
{\bf q}_1 \cdot ({\bf p}_3 -{\bf p}_1)
\Bigr]
\Biggr\}^2   
\eea

Notice that $A_\vee = 0$ and $A_Y = 0$
whenever two of the three momenta are equal or
the three momenta are parallel. $A_Y$ is subleading in the relevant regime and
we can neglect it.

In order to compare $A_\vee$
with matrix theory we consider the {\it eikonal expression}
where we integrate over the time $t$ along the world-line trajectories 
the Fourier transform 
\be
a_\vee =  \int \frac{\di^9{\bf q}_1 \di^9{\bf q}_2}{(2\pi)^{18}} \: A_\vee 
\: \exp \Bigl[ i \: {\bf q}_1 \cdot  ({\bf r}_1 - {\bf r}_3) 
+   i \: {\bf q}_2 \cdot  ({\bf r}_2 - {\bf r}_3)\Bigr] \, ,
\ee
where ${\bf r}_{i} = (v_i {\bf\hat n}_1 t +{\bf b}_{i})$,
${\bf b}_i\cdot{\bf\hat n}_1=0$ and
$B\equiv |{\bf b}_1 -{\bf b}_2| \gg b \equiv |{\bf b}_2 -{\bf b}_3|$.
We write the momenta in terms of the 
velocities as ${\bf p}_i =M_i {\bf v}_i$ while
bearing in mind that $M_i\sim N_i$. 
 We normalize the amplitude by dividing the result
by the product of the $M_i$ and find~\cite{FFI}
\be
\tilde{a}_\vee  \sim  \int \di t\; 
\frac{N_1 N_2 N_3 v_{23}^2 v_{13}^2 v_{12}^2}{(v_{23}^2t^2 + B^2)^{7/2}
(v_{12}^2t^2 + b^2)^{7/2}}  \sim 
\frac{N_1 N_2 N_3 |v_{23}| v_{13}^2 v_{12}^2}{B^7 b^6}
\ee
that is to be compared to matrix theory.

A bit of controversy arised concerning the term $\tilde{a}_\vee$.
It was thought to be impossible 
in matrix theory~\cite{DR}. However, the argument was not correct, as first
shown in~\cite{FFI}.

The matrix theory computation is in this case
 based on the rank two group $SU(3)$.
We choose the background
\be
   R_1 =\pmatrix{v_1 t & 0     & 0     \cr
                 0     & v_2 t & 0     \cr
                 0     & 0     & v_3 t \cr} \qquad\hbox{and}\qquad
   R_k =\pmatrix{b_k^1   & 0       & 0       \cr
                 0       & b_k^2   & 0       \cr
                 0       & 0       & b_k^3   \cr}\quad k>1.
\ee

We can factor out the motion 
of the center of mass by imposing $v_1 + v_2 + v_3 = 0$ and
$b_k^1 + b_k^2 + b_k^3 = 0$. 

We use a Cartan basis for $SU(3)$, where $H^1$ and $H^2$ denote the 
generators of the Cartan sub-algebra and $E_\alpha$ ($\alpha=\pm\alpha^1,
\pm\alpha^2,\pm\alpha^3$) the roots. We also define the space vectors
\be
   {\bf R}^\alpha = \sum_{a=1,2}\alpha^a\Tr \Big(H^a {\bf R}\Big) \, .
\label{lim}
\ee
With the standard choice of $H^a$ and $\alpha$, this definition singles out 
the relative velocities and impact parameters, e.g. 
$ R_1^{\alpha^1} = (v_2 - v_3)t\equiv v^{\alpha^1}t$ plus cyclic 
and, for $k>1$, $ R_k^{\alpha^1} = b_k^2 - b_k^3\equiv b_k^{\alpha^1}$ 
plus cyclic. 
According to the previous section we choose the 
relative distance of the first particle with the other two to be much larger 
than the relative distance of particle two and three, in other words, we set
\be
   |{\bf b}^{\alpha^2}|\approx|{\bf b}^{\alpha^3}|\approx B \gg
   |{\bf b}^{\alpha^1}|\approx b \quad \mbox{and} \quad 
B\, b \gg v \, . \label{regime}
\ee 

The propagators and vertices can be easily worked out from the gauge fixed
action (\ref{action}), with two points worth stressing: 
first, the quadratic part (yielding the propagators) is diagonal in root 
space; second, contrary to the $SU(2)$ case, there are now vertices with 
three massive particles (corresponding to the three different roots). The 
second point is particularly crucial because it is from a diagram 
containing those vertices that we find the supergravity term.

We find twenty real massless bosons and
thirty massive complex bosons. 
We only need consider some of the latter
to construct the diagram. Writing $x_k = x_k^a H^a + x_k^\alpha E_\alpha$, with
$x_k^{-\alpha} = x_k^{\alpha *}$, we define the propagators as
\be
  \langle x_k^{\alpha *}(t_1)x_k^{\alpha}(t_2) \rangle  = 
  \Delta\Big( t_1, t_2 \: \Big|\: (b^{\alpha})^2, 
   v^{\alpha}\Big) \, . 
\ee
As for $x_1$, (the fluctuation 
associated to the background $R_1$), it 
mixes with the field $a_0$ (the fluctuation of the gauge potential). Writing
$x_1^\alpha=z^\alpha+w^\alpha$ and $a_0^\alpha=i(z^\alpha-w^\alpha)$ yields
\bea
   \langle z^{\alpha *}(t_1)z^{\alpha}(t_2) \rangle  &=&
   \Delta\Big(t_1, t_2  \: \Big| \: (b^{\alpha})^2+2v^{\alpha}, 
   v^{\alpha}\Big) \nn \\
   \langle w^{\alpha *}(t_1)w^{\alpha}(t_2) \rangle &=&
   \Delta\Big( t_1, t_2  \: \Big| \: (b^{\alpha})^2-2v^{\alpha}, 
   v^{\alpha}\Big) \, ,
\eea
where
\be
\Delta_i = \int \di s \: e^{-\beta_i^2 s} 
\sqrt{\frac{v^{\alpha^i}}{2\pi\sinh 2\, v^{\alpha^i} s}}\exp\left\{
{-h(v^{\alpha^i}, s)\:t^2
-k(v^{\alpha^i}, s)\:T^2}\right\} 
\ee
where $t=(t_1-t_2)/2$, $T=(t_1+t_2)/2$,
$\beta_1^2 = b^2$, $\beta_2^2 = B^2 + 2 v_{13}$, $\beta_3^2=B^2$ and
\bea
    h(v^{\alpha^i}, s)&=&\frac{v^{\alpha^i}}{\sinh 2\,v^{\alpha^i}s}
                         \Bigl( \cosh 2 \,v^{\alpha^i}s + 1 \Bigr)\nn\\
    k(v^{\alpha^i}, s)&=&\frac{v^{\alpha^i}}{\sinh 2\,v^{\alpha^i}s}
                         \Bigl( \cosh2\,v^{\alpha^i} s - 1 \Bigr)  \, .
\eea

The vertex we need is contained in the term of the effective 
action~(\ref{action}) of type
\be
   -2\:  \Tr \Big( [R_1, x_j][x_1, x_j] \Big) \, ,
\ee
which gives a vertex with two massive bosons and a massless one and another
one with all three massive bosons. Focusing on the second case and choosing a
particular combination of the roots we obtain a term of the type
\be
    v^{\alpha^1}t\; z^{\alpha^2}x_j^{\alpha^1}x_j^{\alpha^3}
    \equiv v_{23}\:t\: z^{13}x_j^{23}x_j^{12} \, , \label{verti}
\ee
and a similar term with $z^{\alpha}$ replaced by $w^{\alpha}$.

The diagrams we have considered
are two-loop  diagrams in the bosonic sector---there are
various similar diagrams which can give rise to the same 
behavior---and we have analyzed in detail one of them, the {\it setting-sun}
diagram with all massive propagators,
which only arises in the three-body problem. 
It can be written as
\be
\tilde a_\ominus = (v^{\alpha^1})^2
\int \di\, t \:\di\, T  \: \left( T^2 - t^2 \right) \Delta_1
\Delta_2\Delta_3  \label{a}
\ee

The appropriate powers of $N_i$ can be 
deduced---following~\cite{BBPT}--- from the double-line notation in which the
setting-sun diagram is of order $N^3$; this factor must be $N_1 N_2 N_3$
for the diagram to involve all three particles.

Expanding (\ref{a}) in the limit (\ref{lim}) yields
\be
\tilde{a}_\ominus \sim 
\frac{N_1 N_2 N_3 |v_{23}| v_{12}^2 v_{13}^2}{B^7b^6}
\ee
which reproduces the behavior of the
supergravity result, that is, 
$\tilde a_\ominus \sim \tilde a_\vee$.

The same result can be obtained in the framework of an effective action in
which the degrees of freedom related to the ``heavy'' modes (those
exchanged at distance $B$) 
are integrated out and the action is discussed in
terms of the ``light'' modes (exchanged at distance $b$). 
Claims about a vanishing
result in such an effective-action approach~\cite{WT} are discussed and
shown to be irrelevant for the three-graviton problem in~\cite{FFI2}.

The preliminary result of ~\cite{FFI} 
concerning a single diagram has been confirmed by
the complete computation performed in~\cite{OY}. They found
perfect numerical agreement for the three graviton scattering in supergravity
and matrix theory.
%
% ---- Bibliography ----
%

\end{document}